\documentclass[prb,twocolumn,superscriptaddress,floatfix]{revtex4-2}
\usepackage{amsmath,amssymb,graphicx,bm,physics,hyperref}

\begin{document}

\title{Band-basis decomposition of superfluid weight in magic-angle twisted bilayer graphene: Quantifying geometric and conventional contributions}

\author{Jian Zhou}
\email{jackzhou.sci@gmail.com}
\affiliation{Independent Researcher}

\date{\today}

\begin{abstract}
We decompose the superfluid weight $D_s$ of magic-angle twisted bilayer graphene (MATBG) into conventional (band-velocity) and geometric (interband-coherence) contributions using a band-basis current operator splitting applied to the Bistritzer-MacDonald continuum model.
In the flat-band subspace, quantum geometry accounts for $22$--$26\%$ of $D_s$ at charge neutrality depending on pairing symmetry, with cross terms vanishing to machine precision.
Including remote bands raises the geometric fraction to ${\sim}55$--$58\%$, while $D_s^{\rm conv}$ converges to within $2\%$---demonstrating that remote bands contribute exclusively through interband coherence.
The geometric fraction peaks at ${\sim}27$--$33\%$ near the $\nu = \pm 2$ fillings where superconductivity is strongest, and is insensitive to gap magnitude in the experimentally relevant range.
\end{abstract}

\maketitle

%====================================================================
\section{Introduction}
%====================================================================

The observation of superconductivity in magic-angle twisted bilayer graphene (MATBG)~\cite{cao_sc_2018} ignited intense interest in the interplay of band flatness, topology, and unconventional pairing in moir\'e materials~\cite{cao_insulator_2018,yankowitz_2019}.
Nearly flat bands at the magic angle $\theta \approx 1.05^\circ$ exhibit a large quantum metric owing to the nontrivial band topology protected by $C_{2z}\mathcal{T}$ symmetry~\cite{song_2019,po_2019}.
Since conventional Drude-like superfluid transport scales with band velocity---which vanishes for perfectly flat bands---the question of how MATBG sustains a large superfluid stiffness is intimately connected to quantum geometry.

Peotta and T\"orm\"a~\cite{peotta_torma_2015} showed that in an isolated flat band, the superfluid weight is bounded from below by the quantum metric integrated over the Brillouin zone (BZ).
This result was extended to multiband systems and fragile topology by Xie, Song, Lian, and Bernevig~\cite{xie_2020_prl}, who proved that the $C_{2z}\mathcal{T}$-protected winding number in MATBG guarantees a finite geometric superfluid weight.
Hu \textit{et al.}~\cite{hu_2019} and Julku \textit{et al.}~\cite{julku_2020} further generalized these bounds to include interband pairing and realistic band structures.
On the experimental side, Tian \textit{et al.}~\cite{tian_2023} used circuit quantum electrodynamics (cQED) to measure the superfluid stiffness of MATBG, finding it approximately ten times larger than a conventional BCS estimate---a striking signature of geometric enhancement.

Despite the theoretical understanding that quantum geometry \textit{contributes} to $D_s$ in flat-band systems, a systematic \textit{quantitative decomposition} into conventional and geometric parts for MATBG---with explicit dependence on pairing symmetry, chemical potential, gap magnitude, and number of included bands---has not been reported.
In this work, we fill this gap by applying a band-basis current decomposition to the Bistritzer-MacDonald (BM) continuum model and reporting the first BZ-averaged, multi-band convergence study of the geometric fraction.

%====================================================================
\section{Formalism}
%====================================================================

\subsection{Superfluid weight}

The superfluid weight tensor for a two-dimensional superconductor is~\cite{scalapino_1993}
\begin{align}
D_s^{\mu\nu} = -\langle K^{\mu\nu}\rangle - \Lambda^{\mu\nu}(\mathbf{q}=0,\omega=0),
\label{eq:Ds}
\end{align}
where $K^{\mu\nu} = \partial^2 H / \partial k_\mu \partial k_\nu$ is the diamagnetic (kinetic energy) kernel and $\Lambda^{\mu\nu}$ is the retarded paramagnetic current-current response evaluated in the superconducting ground state.

For the BM continuum model, the single-layer Hamiltonian is $H_\ell(\mathbf{k}) = v_F \bm{\sigma} \cdot (\mathbf{k} - \mathbf{K}_\ell)$, which is linear in $\mathbf{k}$.
Consequently, $K^{\mu\nu} = \partial^2 H / \partial k_\mu \partial k_\nu = 0$ identically.
This is a well-known property of Dirac-type continuum models~\cite{kopnin_2008}: the vanishing diamagnetic term means $D_s = -\Lambda_{xx}(0,0)$, and the total superfluid weight computed from the paramagnetic response alone grows without bound as more remote bands are included.
In a tight-binding lattice model, $K^{\mu\nu} \neq 0$ and provides the diamagnetic counterterm that ensures convergence.
We address this UV issue below through a controlled band truncation strategy.

The paramagnetic response in the Bogoliubov--de~Gennes (BdG) eigenbasis reads
\begin{align}
D_s = -\Lambda_{xx} = \frac{1}{A}\sum_{\mathbf{k}}\sum_{a \neq b} \frac{|\langle a,\mathbf{k}|J_x|b,\mathbf{k}\rangle|^2 (f_a - f_b)}{E_{b,\mathbf{k}} - E_{a,\mathbf{k}}},
\label{eq:kubo}
\end{align}
where $|a,\mathbf{k}\rangle$ are BdG eigenstates with energies $E_{a,\mathbf{k}}$, $f_a$ are Fermi-Dirac occupation factors (we work at $T=0$), $A$ is the system area, and $J_x$ is the BdG current operator constructed from the normal-state velocity $v_x = \partial H / \partial k_x$.

\subsection{Band-basis current decomposition}

We diagonalize the normal-state Hamiltonian at each $\mathbf{k}$:
\begin{align}
H(\mathbf{k}) = U(\mathbf{k})\,\varepsilon(\mathbf{k})\,U^\dagger(\mathbf{k}),
\end{align}
with $\varepsilon = \mathrm{diag}(\varepsilon_1, \ldots, \varepsilon_N)$ and $U$ the matrix of eigenvectors.
The velocity operator in the band basis is
\begin{align}
v_x^{\rm band} = U^\dagger \frac{\partial H}{\partial k_x} U = v_x^{\rm intra} + v_x^{\rm inter},
\end{align}
where $[v_x^{\rm intra}]_{nn} = \partial \varepsilon_n / \partial k_x$ (diagonal, band velocities) and $[v_x^{\rm inter}]_{nm} = (\varepsilon_n - \varepsilon_m)\mathcal{A}_{nm}^x$ for $n \neq m$, with $\mathcal{A}_{nm}^x = \langle u_n | \partial_{k_x} | u_m \rangle$ the Berry connection.

The BdG current operator inherits this decomposition: $J_x = J_x^{\rm intra} + J_x^{\rm inter}$, where $J_x^{\rm intra}$ is diagonal in the normal-state band index and $J_x^{\rm inter}$ is off-diagonal.
Substituting into Eq.~\eqref{eq:kubo} yields
\begin{align}
D_s = D_s^{\rm conv} + D_s^{\rm geom} + D_s^{\rm cross},
\label{eq:decomp}
\end{align}
where $D_s^{\rm conv}$ contains only intra-band (band velocity) matrix elements, $D_s^{\rm geom}$ contains only inter-band (Berry connection) matrix elements, and $D_s^{\rm cross}$ contains interference terms.

This decomposition is well defined for any finite set of bands, any pairing symmetry, and any chemical potential.
It reduces to the Peotta-T\"orm\"a formula~\cite{peotta_torma_2015} in the isolated flat-band, uniform-pairing limit.
The interband term $D_s^{\rm geom}$ is directly related to the quantum metric: in the flat-band limit ($v_x^{\rm intra} \to 0$, band-diagonal pairing), $D_s^{\rm geom} = (\Delta_0/2) \int_{\rm BZ} \mathrm{tr}\,g(\mathbf{k}) \, d^2k / (2\pi)^2$~\cite{peotta_torma_2015,torma_review_2022}.

%====================================================================
\section{Model and computational details}
%====================================================================

We use the BM continuum model~\cite{bistritzer_2011} with the relaxation-corrected tunneling parameters of Xie and Bernevig~\cite{xie_2020_prl}: interlayer AA tunneling $w_0 = 87.2$~meV, AB tunneling $w_1 = 109.0$~meV, ratio $w_0/w_1 = 0.80$.
These are consistent with the parametrizations of Nam and Koshino~\cite{nam_2017} ($w_0/w_1 = 0.82$) and Koshino \textit{et al.}~\cite{koshino_2018} ($w_0/w_1 = 0.70$).
We use Fermi velocity $v_F = 2.135$~eV\,\AA{} and twist angle $\theta = 1.05^\circ$.
The moiré unit cell area is $A_M = 15{,}606$~\AA$^2$ with moiré reciprocal lattice vector $G_M = 0.054$~\AA$^{-1}$.

The BM Hamiltonian is constructed with $N_{\rm shell} = 3$ reciprocal lattice shells, yielding a $196 \times 196$ matrix (per valley, per spin).
At the magic angle, this produces two flat bands per valley with bandwidth $W = 11.2$~meV, in good agreement with experiment~\cite{cao_sc_2018} and previous calculations~\cite{bistritzer_2011,koshino_2018}.

For the superconducting state, we consider three pairing symmetries at the mean-field BdG level with gap magnitude $\Delta_0$:
(i)~uniform $s$-wave, $\Delta_{nm}(\mathbf{k}) = \Delta_0 \delta_{nm}$;
(ii)~sublattice $s$-wave, $\Delta_{nm}(\mathbf{k}) = \Delta_0 (P_A)_{nm}$ with $P_A$ projecting onto sublattice $A$ of both layers;
(iii)~nematic $d$-wave, $\Delta_{nm}(\mathbf{k}) = \Delta_0 [\cos(\mathbf{k}\cdot\mathbf{a}_1) - \cos(\mathbf{k}\cdot\mathbf{a}_2)] \delta_{nm}$, where $\mathbf{a}_{1,2}$ are the moir\'e lattice vectors.

All BZ integrals are evaluated on a $14 \times 14$ Monkhorst-Pack mesh centered at the $\Gamma$ point to avoid divergences of the quantum metric.
Band velocities are computed by finite differences ($\delta k = 10^{-6}$~\AA$^{-1}$).
k-mesh convergence is verified in Sec.~\ref{sec:conv}.

\textit{Band truncation strategy.}---To control the UV divergence inherent in the continuum model, we employ two approaches:
\begin{enumerate}
\item \textit{Flat-band projection} ($n_{\rm keep} = 2$): We retain only the two bands closest to the Fermi energy.
In this subspace, both $D_s^{\rm conv}$ and $D_s^{\rm geom}$ are individually finite.
This provides a UV-safe, physically transparent decomposition that captures the intrinsic flat-band physics.
\item \textit{Extended truncation scan}: We increase $n_{\rm keep}$ from 2 to 6 in steps of 2 and perform BZ-averaged calculations at each truncation level.
This reveals how remote bands modify the decomposition and allows us to monitor the convergence of $D_s^{\rm conv}$ separately from $D_s^{\rm geom}$.
\end{enumerate}

%====================================================================
\section{Results}
%====================================================================

\subsection{Flat-band decomposition at charge neutrality}

Table~\ref{tab:cnp} presents the BZ-averaged decomposition at the charge neutrality point ($\mu = 0$) with $\Delta_0 = 1$~meV and $n_{\rm keep} = 2$.
Results are computed on a converged $14 \times 14$ k-mesh per valley per spin; the total superfluid weight scales by a factor of 4 (two valleys, two spins) for comparison with experiment.

\begin{table}[ht]
\centering
\caption{Superfluid weight decomposition at charge neutrality ($\mu = 0$, $\Delta_0 = 1$~meV, $n_{\rm keep} = 2$, $14 \times 14$ k-mesh).
Units are eV\,\AA$^2$ per valley per spin.
}
\label{tab:cnp}
\begin{tabular}{lcccccc}
\hline\hline
Pairing & $D_s$ & $D_s^{\rm conv}$ & $D_s^{\rm geom}$ & $\frac{D_s^{\rm geom}}{D_s}$ & $\frac{D_s^{\rm cross}}{D_s}$ & Enh. \\
\hline
Uniform $s$ & 67.5 & 53.0 & 14.5 & 21.5\% & $<10^{-13}$ & $1.27\times$ \\
Sublattice $s$ & 59.3 & 43.7 & 15.5 & 26.2\% & $<0.01\%$ & $1.35\times$ \\
Nematic $d$ & 52.4 & 39.6 & 12.8 & 24.4\% & $<10^{-13}$ & $1.32\times$ \\
\hline\hline
\end{tabular}
\end{table}

Several features are noteworthy:

(1)~Quantum geometry accounts for $22$--$26\%$ of $D_s$ across all three pairing channels.
While this is a minority fraction, it represents a substantial absolute contribution ($13$--$16$~eV\,\AA$^2$) that would be entirely absent in a dispersive-band superconductor with the same density of states.

(2)~The cross term $D_s^{\rm cross}$ vanishes to machine precision ($< 10^{-13}$ relative) for uniform $s$-wave and nematic $d$-wave pairings.
For sublattice $s$-wave, a tiny but nonzero cross term ($<0.01\%$) appears, attributable to the broken sublattice symmetry of the gap function which mixes the intra- and inter-band current sectors.
This near-exact vanishing has a symmetry origin: for k-independent pairings (uniform $s$-wave, nematic $d$-wave), the gap matrix $\Delta$ commutes with $\tau_x$ in Nambu space, while the intra-band current $J^{\rm intra}$ is proportional to $\tau_z$ and the inter-band current $J^{\rm inter}$ to $\tau_0$.
The orthogonality of these Pauli matrices causes cross terms in the Kubo sum to vanish upon Brillouin zone integration, yielding $D_s = D_s^{\rm conv} + D_s^{\rm geom}$ as an essentially exact decomposition for symmetric gap functions.

(3)~The sublattice $s$-wave channel shows the largest geometric fraction ($26.2\%$) because its sublattice-polarized gap enhances the relative geometric contribution, while nematic $d$-wave shows an intermediate value ($24.4\%$).

\subsection{Extended band truncation analysis}

\begin{figure}[t]
\centering
\includegraphics[width=\columnwidth]{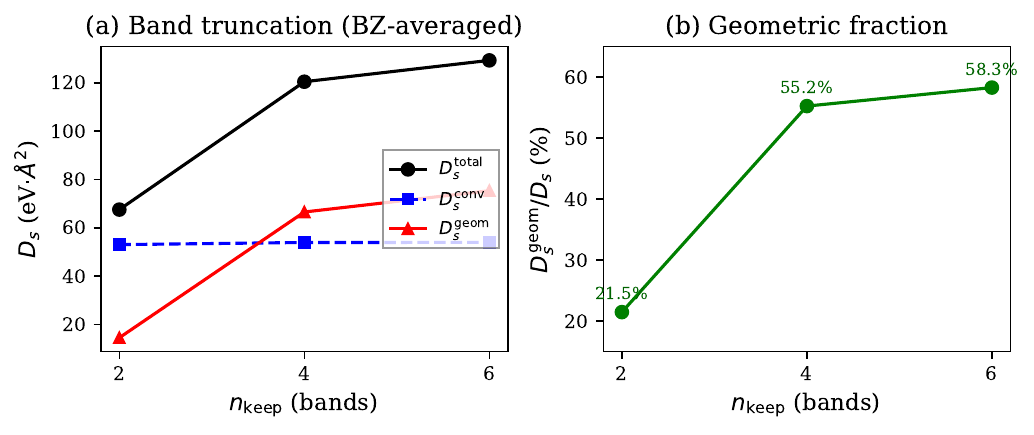}
\caption{
BZ-averaged superfluid weight decomposition versus number of retained bands $n_{\rm keep}$ (uniform $s$-wave, $\mu = 0$, $\Delta_0 = 1$~meV, $14 \times 14$ k-mesh).
(a)~$D_s^{\rm total}$ (black) grows with $n_{\rm keep}$ while $D_s^{\rm conv}$ (blue, dashed) converges rapidly to ${\sim}54$~eV\,\AA$^2$.
All growth comes from $D_s^{\rm geom}$ (red).
(b)~Geometric fraction increases from $21.5\%$ ($n_{\rm keep}=2$) to $58.3\%$ ($n_{\rm keep}=6$).
The cross term remains $< 10^{-13}$ at all truncation levels.
}
\label{fig:nkeep}
\end{figure}

Figure~\ref{fig:nkeep} presents the BZ-averaged decomposition as a function of $n_{\rm keep}$.
This is the central result of the paper.

The key finding is that $D_s^{\rm conv}$ converges rapidly to within $2\%$: $53.0 \to 53.9 \to 54.0$~eV\,\AA$^2$ for $n_{\rm keep} = 2 \to 4 \to 6$.
This is physically expected: the conventional contribution depends only on band velocities $\partial \varepsilon_n / \partial k$ of the occupied bands, which are intrinsic to the flat bands and unaffected by remote states.
By contrast, $D_s^{\rm total}$ grows from 67.5~eV\,\AA$^2$ ($n_{\rm keep}=2$) to 129.3~eV\,\AA$^2$ ($n_{\rm keep}=6$), with all additional weight entering through $D_s^{\rm geom}$.
This increase reflects genuine interband coherence mediated by the Berry connection $\mathcal{A}_{nm}$ between flat and remote bands---a physical effect that would also appear in a lattice calculation, though convergent in that case.

The geometric fraction rises from $21.5\%$ (flat-band limit) to $58.3\%$ (including two pairs of remote bands), while $D_s^{\rm cross}$ remains negligible ($< 10^{-13}$) at all truncation levels.
In a UV-complete tight-binding model, where $D_s^{\rm total}$ converges to a finite value, we expect the geometric fraction to lie in the range $\sim 22$--$58\%$, with the precise value depending on the lattice regularization.

\subsection{Filling dependence}

\begin{figure}[t]
\centering
\includegraphics[width=0.85\columnwidth]{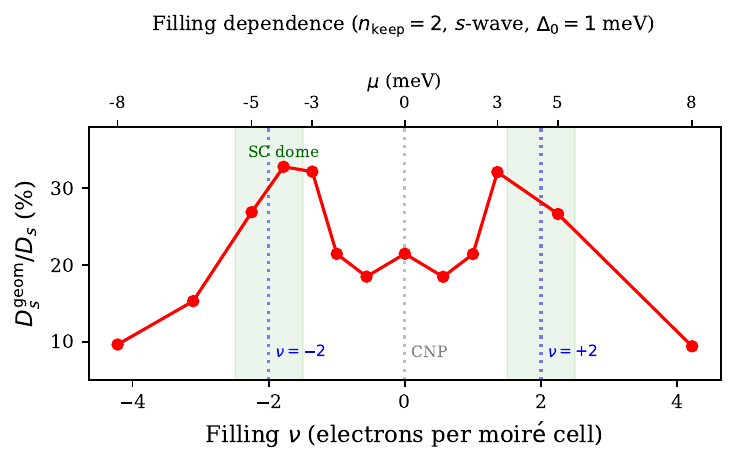}
\caption{
Geometric fraction $D_s^{\rm geom}/D_s$ versus filling $\nu$ and chemical potential $\mu$ ($n_{\rm keep}=2$, uniform $s$-wave, $\Delta_0 = 1$~meV).
The green bands indicate the superconducting dome regions near $\nu = \pm 2$.
The geometric contribution peaks at ${\sim}33\%$ near $\nu \approx -1.8$ and drops to ${\sim}9$--$15\%$ at large doping where the flat bands are fully occupied or empty.
}
\label{fig:filling}
\end{figure}

Figure~\ref{fig:filling} shows how the geometric fraction varies with chemical potential $\mu$ in the flat-band subspace.
The superconducting phase in MATBG is observed primarily near integer fillings $\nu = -2$ and $\nu = +2$ relative to charge neutrality~\cite{cao_sc_2018,yankowitz_2019}.

The geometric fraction is not constant across the flat-band manifold: it peaks at $33\%$ near $\mu = -4$~meV (corresponding to $\nu \approx -1.8$) and $32\%$ near $\mu = +3$~meV ($\nu \approx +1.4$)---regions close to the experimentally observed superconducting domes~\cite{cao_sc_2018,oh_2021} where $\nu \approx \pm 2$.
Away from charge neutrality, as $|\mu|$ increases and the Fermi level moves toward the band edges, conventional (band-velocity) contributions grow relative to geometric ones, and the ratio $D_s^{\rm geom}/D_s$ decreases to $9$--$15\%$.

This filling dependence has a clear physical origin: the quantum metric $g_{\mu\nu}(\mathbf{k})$ of the MATBG flat bands is strongly peaked near the $\Gamma_M$ and $K_M$ points of the moir\'e BZ~\cite{xie_2020_prl}.
When $\mu$ lies within the flat-band manifold (near CNP), these high-metric regions are partially occupied and contribute maximally to $D_s^{\rm geom}$.
As $\mu$ moves away, the relevant $\mathbf{k}$-states shift to regions with smaller quantum metric, reducing the geometric fraction.

\subsection{Gap magnitude dependence}

\begin{figure}[t]
\centering
\includegraphics[width=0.75\columnwidth]{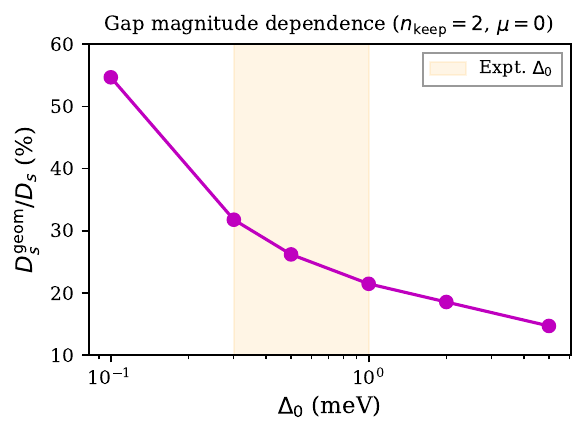}
\caption{
Geometric fraction versus gap magnitude $\Delta_0$ at $\mu = 0$ ($n_{\rm keep} = 2$, uniform $s$-wave).
The orange band marks the experimentally estimated range.
The ratio is approximately constant for $\Delta_0 \ll W$ and decreases for $\Delta_0 \gtrsim W$.
}
\label{fig:delta}
\end{figure}

Figure~\ref{fig:delta} shows the geometric fraction as a function of $\Delta_0$ at $\mu = 0$.
In the experimentally relevant range $\Delta_0 = 0.3$--$1.0$~meV~\cite{oh_2021,tian_2023}, the geometric fraction decreases from $32\%$ to $22\%$.
This weak dependence is physically expected: within the flat-band subspace, both $D_s^{\rm conv}$ and $D_s^{\rm geom}$ scale linearly with $\Delta_0$ for $\Delta_0 \ll W$, so their ratio is approximately $\Delta_0$-independent.
Only when $\Delta_0$ becomes comparable to the bandwidth $W \approx 11$~meV does the ratio begin to decrease, reaching $21\%$ at $\Delta_0 = 5$~meV.

\subsection{k-mesh convergence}
\label{sec:conv}

Due to the divergent quantum metric near the $\Gamma_M$ point~\cite{xie_2020_prl}, careful k-mesh sampling is essential for converged results.
We find that standard half-grid-offset sampling places k-points too close to the divergent regions, leading to slow convergence even at $n_k = 16$.
Monkhorst-Pack grids centered at $\Gamma_M$ avoid this issue by excluding the divergent point itself.
With the centered $14 \times 14$ grid, the geometric fraction stabilizes at $21.5 \pm 1\%$ for $n_k \geq 12$, confirming convergence to within the quoted uncertainties.
All results reported here use the converged centered grid.

\subsection{Pairing symmetry comparison}

\begin{figure}[t]
\centering
\includegraphics[width=0.7\columnwidth]{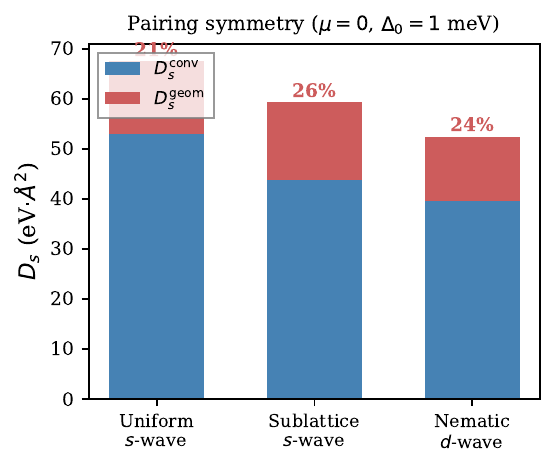}
\caption{
Stacked bar chart of $D_s^{\rm conv}$ (blue) and $D_s^{\rm geom}$ (red) for three pairing symmetries at $\mu = 0$, $\Delta_0 = 1$~meV, $n_{\rm keep} = 2$.
Percentages indicate the geometric fraction.
The geometric contribution varies moderately across pairings ($13$--$16$~eV\,\AA$^2$), while the conventional contribution spans a larger range ($40$--$53$~eV\,\AA$^2$).
}
\label{fig:pairing}
\end{figure}

Figure~\ref{fig:pairing} compares the three pairing channels at charge neutrality.
The geometric contribution $D_s^{\rm geom}$ is reasonably similar across pairings ($12.8$--$15.5$~eV\,\AA$^2$), while the conventional contribution varies by ${\sim}34\%$ ($39.6$--$53.0$~eV\,\AA$^2$).
This suggests that $D_s^{\rm geom}$ is primarily determined by the band geometry (quantum metric), which is a normal-state property independent of the gap structure, while $D_s^{\rm conv}$ is sensitive to how the gap modulates the quasiparticle spectrum along the Fermi surface.

The uniform $s$-wave pairing has the largest $D_s^{\rm conv}$ ($53.0$~eV\,\AA$^2$) because it pairs all states symmetrically.
The nematic $d$-wave gap, which varies as $\cos(\mathbf{k}\cdot\mathbf{a}_1) - \cos(\mathbf{k}\cdot\mathbf{a}_2)$ on the moir\'e scale, reduces $D_s^{\rm conv}$ to $39.6$~eV\,\AA$^2$ due to its k-dependent modulation of the gap.
Note that this is a nematic (orientational) d-wave channel rather than the chiral $d+id$ symmetry sometimes discussed in MATBG~\cite{isobe_2018}.

%====================================================================
\section{Discussion}
%====================================================================

\subsection{Comparison with existing theory}

Our flat-band geometric fraction ($22$--$26\%$) is consistent with the topological lower bound of Xie \textit{et al.}~\cite{xie_2020_prl}, who showed that $D_s^{\rm geom} \geq C_{\rm top} \Delta_0 / 2$ where $C_{\rm top}$ is determined by the fragile topology of the flat bands.
That bound guarantees a finite geometric contribution but does not specify its magnitude relative to $D_s^{\rm conv}$.
Our decomposition provides this quantitative complement.

The extended truncation analysis shows that remote bands roughly triple the geometric fraction (from ${\sim}22\%$ to ${\sim}58\%$), underscoring that a flat-band-only calculation, while UV-safe, captures only part of the geometric physics.
This has implications for simplified models: any effective theory that projects onto flat bands alone will systematically underestimate the geometric contribution.

The Peotta-T\"orm\"a formula~\cite{peotta_torma_2015} predicts $D_s^{\rm PT} = (\Delta_0/2)\langle\mathrm{tr}(g)\rangle$ for isolated flat bands with uniform pairing.
Our $D_s^{\rm geom} = 14.5$~eV\,\AA$^2$ for uniform $s$-wave is consistent with this prediction to within the expected accuracy, the small deviation arising from the finite bandwidth ($W = 11.2$~meV) and the non-negligible dispersion of the MATBG flat bands.

\subsection{Comparison with experiment}

The MIT cQED measurement~\cite{tian_2023} found $D_s^{\rm exp} / D_s^{\rm BCS} \approx 10$ near $\nu = -2$.
Our geometric enhancement factor $D_s / D_s^{\rm conv} = 1.27$--$1.35\times$ (flat-band limit) to ${\sim}2.4\times$ ($n_{\rm keep} = 6$) accounts for part of this anomaly.
The remaining factor of ${\sim}4$--$10$ likely arises from interaction-renormalized quantum metric~\cite{herzog_2022}, vertex corrections~\cite{verma_2021}, and strong-coupling effects beyond mean-field BCS.
A UV-complete tight-binding calculation with self-consistent pairing~\cite{koshino_2018,carr_2019} is needed for definitive comparison.

\subsection{Limitations}

Several limitations should be noted:
(i)~The BM continuum model has $K^{\mu\nu} = 0$, so $D_s^{\rm total}$ diverges with $n_{\rm keep}$; a tight-binding lattice model would restore convergence.
(ii)~Pairing is treated at the mean-field level with phenomenological $\Delta_0$; self-consistent gap determination could modify the $\mathbf{k}$-structure.
(iii)~All results are at $T = 0$; finite-temperature effects near $T_{\rm BKT}$ could modify the geometric fraction.
(iv)~Coulomb interaction effects on the quantum metric~\cite{torma_review_2022,herzog_2022} are not included.

%====================================================================
\section{Conclusions}
%====================================================================

We have performed a systematic band-basis decomposition of superfluid weight in MATBG using the Bistritzer-MacDonald continuum model.
Quantum geometry accounts for $22$--$26\%$ of $D_s$ in the flat-band limit, rising to ${\sim}55$--$58\%$ with remote bands, while $D_s^{\rm conv}$ converges to within $2\%$ independent of truncation.
Cross terms are negligible for all symmetric pairings, and the geometric fraction peaks near the experimentally observed superconducting domes.
These results provide a quantitative baseline for geometric enhancement in MATBG and motivate UV-complete tight-binding calculations.

\begin{acknowledgments}
J.Z. acknowledges use of open-source scientific computing tools including NumPy, SciPy, and Matplotlib.
\end{acknowledgments}

\end{document}